\documentclass[conference]{IEEEtran}

\usepackage{orcidlink}
\usepackage{booktabs}
\usepackage{multirow}
\usepackage[ruled, lined, linesnumbered, commentsnumbered, longend]{algorithm2e}

\usepackage{tcolorbox}

\usepackage{adjustbox}

\usepackage{cite}
\usepackage{amsmath,amssymb,amsfonts}
\usepackage{algorithmic}
\usepackage{graphicx}
\usepackage{textcomp}
\usepackage{xcolor}
\usepackage{enumitem}
\def\BibTeX{{\rm B\kern-.05em{\sc i\kern-.025em b}\kern-.08em
    T\kern-.1667em\lower.7ex\hbox{E}\kern-.125emX}}

\usepackage{amsthm}

\newcommand\copyrighttext{\footnotesize \textcopyright 2023 IEEE. Personal use of this material is permitted. Permission from IEEE must be obtained for all other uses, in any current or future media, including reprinting/republishing this material for advertising or promotional purposes, creating new collective works, for resale or redistribution to servers or lists, or reuse of any copyrighted component of this work in other works.}
\newcommand\copyrightnotice{%
\begin{tikzpicture}[remember picture,overlay]
\node[anchor=south,yshift=10pt] at (current page.south) {\fbox{\parbox{\dimexpr\textwidth-\fboxsep-\fboxrule\relax}{\copyrighttext}}};
\end{tikzpicture}%
}

\theoremstyle{definition}
\newtheorem{definition}{Definition}[section]    

\begin{document}
\bstctlcite{bstctl:nodash}

\title{Property-Based Mutation Testing}
\author{
\IEEEauthorblockN{Ezio Bartocci}
\IEEEauthorblockA{
\textit{TU Wien}\\
Vienna, Austria \\
ezio.bartocci@tuwien.ac.at}
\and
\IEEEauthorblockN{Leonardo Mariani}
\IEEEauthorblockA{
\textit{University of Milano-Bicocca}\\
Milan, Italy \\
leonardo.mariani@unimib.it}
\and
\IEEEauthorblockN{Dejan Ničković}
\IEEEauthorblockA{
\textit{Austrian Institute of Technology}\\
Vienna, Austria \\
Dejan.Nickovic@ait.ac.at}
\and
\IEEEauthorblockN{Drishti Yadav}
\IEEEauthorblockA{
\textit{TU Wien}\\
Vienna, Austria \\
drishti.yadav@tuwien.ac.at}
}

\maketitle
\copyrightnotice

\newcommand{\changes}[1]{\textcolor{black}{#1}}

\begin{abstract}

Mutation testing is an established software quality assurance technique for the assessment of test suites. While it is well-suited to estimate the general fault-revealing capability of a test suite, it is not practical and informative when the software under test must be validated against specific requirements. This is often the case for embedded software, where the software is typically validated against rigorously-specified safety properties. In such a scenario (i) a mutant is relevant only if it can impact the satisfaction of the tested properties, and (ii) a mutant is meaningfully-killed with respect to a property only if it causes the violation of that property. 
To address these limitations of mutation testing, we introduce  \emph{property-based mutation testing}, a method for assessing the capability of a test suite to exercise the software with respect to a given property. We evaluate our property-based mutation testing framework on Simulink models of safety-critical Cyber-Physical Systems (CPS) from the automotive and avionic domains and demonstrate how property-based mutation testing is more informative than regular mutation testing. These results open new perspectives in both mutation testing and test case generation of CPS.   

\end{abstract}


\section{Introduction}\label{sec1}
Software has a pivotal role in safety-critical applications, from autonomous vehicles to medical devices. Inadequate software quality assurance may result in potentially catastrophic system failures. It is thus important to thoroughly test software, checking that it does not violate its critical properties. 

Mutation testing (MT) is a well-established technique to measure the adequacy of a test suite w.r.t. a fault model~\cite{DeMilloLS78,AcreeBDLS79,olivares2021mutation,fortunato2022mutation}: MT first injects some artificial defects in the software-under-test, and then 
measures the thoroughness of the test suite as the percentage of injected faults that the test suite can reveal. The injection is performed through \emph{mutation operators} that modify the software according to well-defined patterns. The resulting modified program is called a \textit{mutant}. A test case \emph{kills} a mutant if its execution causes observable differences in the behavior of the original and mutated programs. The ratio of killed mutants w.r.t. the mutants that are not equivalent to the original program is known as the \emph{mutation score}. Ideally, a test suite should reach a mutation score equal to one.


While MT is effective when the test suite has to be assessed against a wide set of faults spread in the software, it loses its effectiveness when the purpose of a test suite is to validate the software against specific requirements. This is particularly true in the embedded software domain, where software must be often validated against rigorously-defined safety properties. For example, the \textit{ATCS} (Automatic Transmission Controller System) we used in the experimental evaluation is annotated with several safety properties expressed with Signal Temporal Logic (STL)~\cite{maler2013monitoring}, and test cases are designed to validate the software against these properties. 

When applying mutation testing to assess the capability of a test suite to thoroughly exercise a software \changes{w.r.t.} 
a given property, there are two challenges to take into consideration: the relevance of the mutants and the relevance of the executions that kill the mutants. 

\emph{Relevance of the mutants \changes{w.r.t.} 
a tested property}. Not all the mutants are relevant to assess the thoroughness of a test suite against a property. In fact, only the mutants whose effects propagate in a way that ultimately causes the property violation 
are relevant. A mutant that does not impact a property shall also not contribute to measuring the adequacy of a test suite against that property. 
Regular MT does not distinguish between these mutants, and hence does not consider 
the difference between them when computing the mutation score. 

\begin{figure}[htbp]
\centerline{\includegraphics[width=0.9\linewidth]{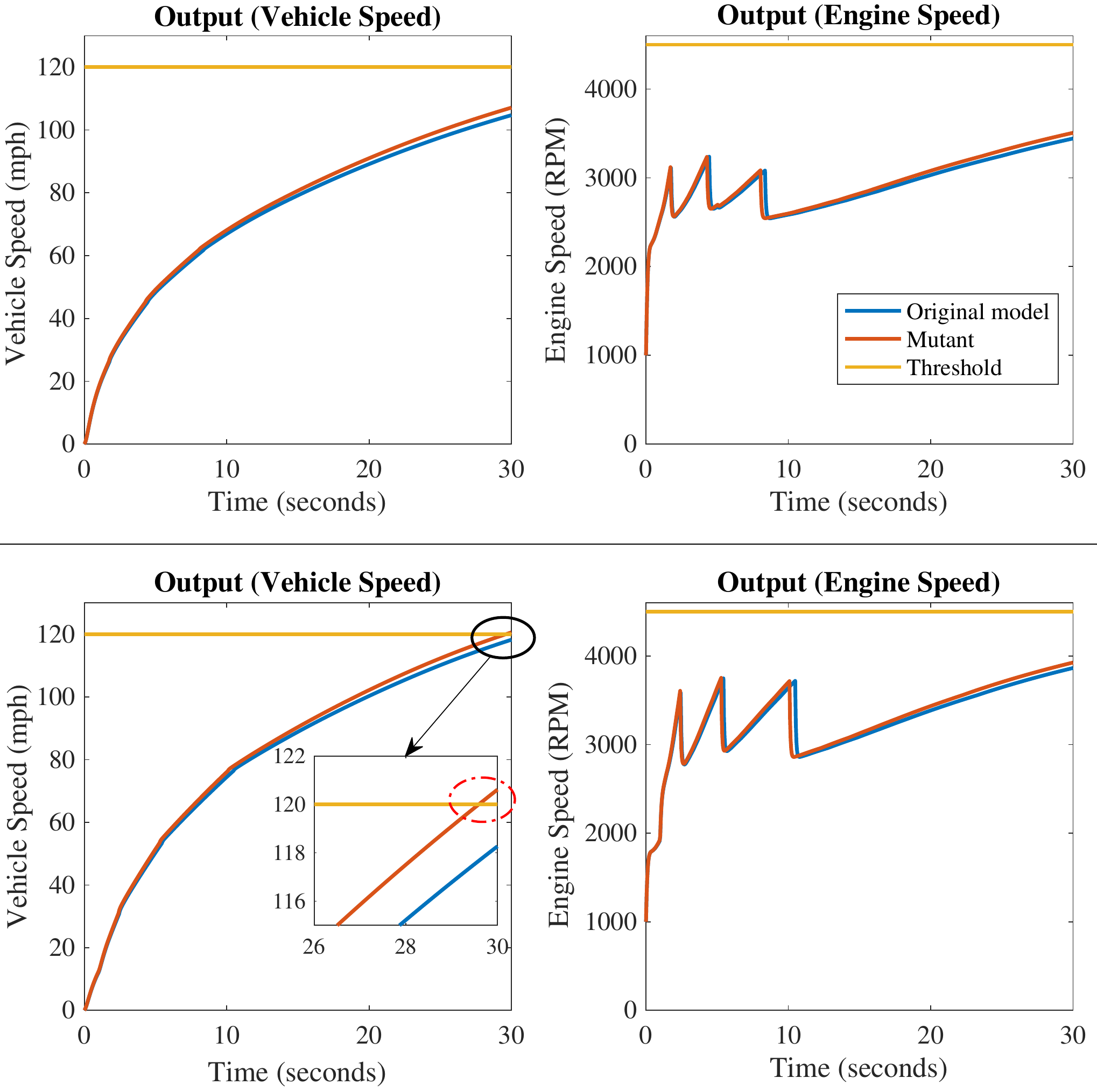}}
\caption{Output plots for the original and mutated models of ATCS: (\textbf{top}) for a test case satisfying the property on the mutant, (\textbf{bottom}) for a test case violating the property on the mutant. The portion of the output trace (vehicle speed) responsible for property violation is highlighted.}
\label{fig:1}
\end{figure}

\emph{Relevance of the execution that kills a mutant}. Producing different outputs for the original and the mutated programs is insufficient to kill a mutant when a test suite is assessed against a property. In fact, a test is thoroughly exercising the software \changes{w.r.t.} a property only if the difference in the two outputs is severe and relevant enough to cause a violation of the property under consideration. Otherwise, the test is generating differences that are marginal \changes{w.r.t.} the testing objective. For instance, in our evaluation, we assessed the test cases for the ATCS against the property that requires the engine speed and the vehicle speed to remain below certain thresholds. Several tests succeeded in exercising a mutant in the Transmission component, causing differences in the outputs, but failed to produce outputs that violate these properties, which is a clear inadequacy of the test suite. This situation is visually illustrated in Fig.~\ref{fig:1} (top)
, where the test is generating differences in the engine and vehicle speeds without exceeding the threshold. The mutant would be counted as killed according to regular mutation testing, although the test does not make the software to violate the property. In practice, if the fault would be present in the original model, the test would not reveal it. This also exemplifies how mutations could be easily killed according to regular mutation testing in data-flow models, where most of the components are activated in every computation and values easily propagate through the blocks in the model. However, the propagated values often result in minor and non-significant output differences. Killing mutants while taking the tested properties under consideration is a definitely harder challenge. For instance, Fig.~\ref{fig:1} (bottom) shows the case of a test that reveal the mutant by violating the tested properties, obtained in our experiments.

In this paper, we address these challenges by defining the notion of \textit{Property-Based Mutation Testing (PBMT)} 
to assess test suites against properties or specifications. To this end, we revise the key notions of mutation testing to measure the effectiveness of the test suites as their capability to exercise the software against a property. We also define a search-based test generation strategy for Simulink models to effectively and automatically identify the relevant mutants that could be killed with meaningful executions, from a set of injected mutants. We provide empirical evidence that PBMT is more informative than MT to assess the thoroughness of test suites, considering two benchmarks in the domain of safety-critical CPS with requirements expressed in STL formalism.



In summary, this paper makes the following contributions:
\begin{enumerate}
    \item We introduce the novel notion of \emph{Property-Based Mutation Testing} for testing software against properties. 
    \item We define a \emph{search-based strategy} to automatically identify the mutants that contribute to PBMT experiments. 
    \item We report \emph{empirical results} for Simulink models, demonstrating that PBMT is more informative than regular \changes{MT} when software is tested against properties.  
    \item We make \emph{tools and experimental data} publicly available for reproduction and to ease follow-up research\footnote{\changes{\url{https://gitlab.com/DrishtiYadav/mt}}}. 
\end{enumerate}


\emph{Paper Organization}. Section~\ref{sec2} presents the 
overview of regular mutation testing
. Section~\ref{sec3} presents property-based mutation testing, our proposed approach. Section~\ref{sec4} describes testing CPS Simulink models against 
STL specifications. Section~\ref{sec5} presents our evaluation of two safety-critical industrial benchmarks. Section~\ref{sec7} discusses threats to validity. Section~\ref{sec6} describes the lessons learned. 
Section~\ref{sec8} presents related work. Section~\ref{sec9} concludes the paper.




\section{Mutation Testing}\label{sec2}
In this section, we present the background and fundamental concepts of regular mutation testing. 

Mutation testing relies on two fundamental assumptions~\cite{AcreeBDLS79,DeMilloLS78}: (1) the \emph{Competent Programmer Hypothesis} that states that programmers create programs that differ from the correct one mostly by small syntactic errors, and (2) the \emph{Coupling Effect} that asserts that \changes{``complex faults are coupled
to simple faults in such a way that a test data set that detects
all simple faults in a program will detect a high percentage
of the complex faults''~\cite{7927962}.}
Several studies investigate these hypotheses demonstrating that results obtained with mutation testing can reliably predict the results obtained for the vast majority of high-priority real bugs~\cite{Petrovic:Mutants:ICSE:2021,Just:Mutants:FSE:2014,Andrews:Mutantion:TSE:2006, Andrews:Mutants:ICSE:2005}. Although not every bug couples with mutants, mutation testing can still be considered a good tool to measure test suite quality.



We now introduce the key concepts of mutation testing. 

\begin{definition}[\textit{Mutation operator}]
A \emph{mutation operator} is a source-code transformation that introduces a modification in the program-under-test. More rigorously, given a program $\mathcal{P}$, a mutation operator $op$ is a function that takes as inputs $\mathcal{P}$ and a location $k$ inside $\mathcal{P}$ and creates a syntactic alteration of $\mathcal{P}$ at location $k$, if the location can be mutated with $op$.
\end{definition}


\begin{definition}[\textit{Mutant}] For a given program $\mathcal{P}$ and a set of mutation operators $\mathcal{O}$ = \{$op_1, op_2, ..., op_n$\}, a \emph{mutant} $p$ is the result of the application of a mutation operator $op \in \mathcal{O}$ to $\mathcal{P}$ at a specified location $k$. A mutant created by the application of only one mutation operator to $\mathcal{P}$ is known as \textit{First Order Mutant} (FOM). The application of multiple mutation operators to $\mathcal{P}$ results in a \textit{Higher Order Mutant} (HOM)\changes{~\cite{jia2009higher}}.
\end{definition}
    
Given a test suite $\mathcal{T}$, and a test $t \in \mathcal{T}$, we write  $t \models p$ when the test passes on 
$p$ and  $t \not\models p$ when the test fails on 
$p$. We denote with $O(t,p)$ the output generated by $p$ with $t$ and with $\mathcal{T}_U^p$ the (universal) set of every possible valid test case for $p$. 


\begin{definition}[\textit{Killed Mutant}] A mutant $p$ is said to be \emph{killed} by $\mathcal{T}$ if at least one test case $t$ in $\mathcal{T}$ fails when exercising $p$, i.e., $\exists t \in \mathcal{T}: t \not\models p$. 
\end{definition}

\begin{definition}
[\textit{Live Mutant}] Mutants that do not lead to the failure of any test case $t \in \mathcal{T}$ are said to be \emph{live} or survived. Formally, 
$p$ is said to be live if $\forall t \in \mathcal{T},\, t \models p$.
\end{definition}

\begin{definition}
[\textit{Equivalent Mutant}] A mutant $p$ is \emph{equivalent} to the original program $\mathcal{P}$ if they both generate the same output for any possible input. Formally, $p$ is equivalent to $\mathcal{P}$ if $\forall t \in \mathcal{T}_U^\mathcal{P}, O(t,p)=O(t,\mathcal{P})$. 
In other words, no test case can distinguish an equivalent mutant from the original program~\cite{demeyer2020formal}. Note that the detection of equivalent mutants is undecidable. 
\end{definition}

\begin{definition}[\textit{Invalid Mutant}]
A mutant $p$ is considered \emph{invalid} if it cannot be compiled\changes{~\cite{vercammen2022mutation}}. Such a mutant is not included in the mutation coverage.
\end{definition}

\begin{definition}[\textit{Mutation coverage}]
The adequacy of a test suite $\mathcal{T}$  can be measured using the \emph{mutation coverage} (hereafter, \textit{mutation score $\mathcal{MS}$}): the ratio of mutants killed w.r.t. the total number of non-equivalent and valid mutants:
\[\textit{Mutation}\ \textit{coverage} = \frac{\# \textit{killed}\ \textit{mutants}}{\#\textit{valid}\ \textit{mutants}-\# \textit{equivalent}\ \textit{mutants}}\]
$\mathcal{T}$ is said to achieve 100\% mutation test adequacy if it kills all non-equivalent valid mutants. Full mutation coverage ensures that $\mathcal{T}$ is (i) robust against the modeled mutation types, and (ii) sensitive to small changes in the program-under-test ($\mathcal{P}$).

\end{definition}

\begin{definition}
[\textit{Redundant Mutant}] 
\changes{Redundant mutants are not beneficial as they consume resources without contributing to the test process as they are killed whenever other mutants are killed.}
This redundancy can be expressed by \textit{duplicate} and \textit{subsumed} mutants~\cite{papadakis2019mutation}. \emph{Duplicate} mutants are equivalent with each other but not equivalent to the original program~\cite{olivares2021mutation}. \emph{Subsumed} mutants are not equivalent with each other but are killed by the same test cases. The subsumption relation is defined as follows~\cite{kurtz2015static}: We say that $p_i$ \textit{subsumes}  $p_j$, denoted $p_i \rightarrow p_j$, iff the following two properties hold:
   \begin{enumerate}
       \item $\exists t \in \mathcal{T}_U^\mathcal{P}: t \not\models p_i$. In other words, there exists some test case $t$ s.t. $p_i$ and $\mathcal{P}$ yield different outputs on $t$, i.e., $p_i$ is not equivalent to $\mathcal{P}$.
       \item $\forall t \in \mathcal{T}_U^\mathcal{P}$, if $t \not\models p_i$, then $t \not\models p_j$. In other words, for every possible test case $t$ on $\mathcal{P}$, if $p_i$ yields a different output than $\mathcal{P}$ on $t$, then so does $p_j$.
   \end{enumerate}
\end{definition}

With Regular MT, for a test case $t \in \mathcal{T}$ to kill a mutant $p$, the following three conditions must be satisfied~\cite{offutt1997automatically,offutt2001mutation}:

\begin{enumerate}[leftmargin=*]
    \item \emph{Reachability}: $t$ must \textit{reach} the mutated statement in $p$.
    \item \emph{Necessity}: $t$ must \textit{infect} the program state by causing different program states for $p$ and $\mathcal{P}$.
    \item \emph{Sufficiency}: the incorrect program state must \textit{propagate} to the output of $p$ and be checked by $t$, i.e., there is an observable difference in the outputs of $p$ and $\mathcal{P}$ for $t$.
\end{enumerate}

The above three conditions are known as the \textit{RIP model}. The capability of a test case $t \in \mathcal{T}$ to kill a mutant $p$ is governed by the observability of the program state, leading to following two common types of mutation testing:

\begin{enumerate}[leftmargin=*]
    \item \emph{Weak mutation testing}: A mutant $p$ is killed by a test suite $\mathcal{T}$ if only the first two conditions of the \textit{RIP model} are satisfied. 
    \item \emph{Strong mutation testing}: For a test case $t \in \mathcal{T}$ to kill a mutant $p$, all three conditions of the \textit{RIP model} must be met. 
\end{enumerate}

Tests, in particular automated tests, usually include an explicit comparison of the observed program behavior to the expected behavior using an \textit{oracle}. Thus, automated tests usually examine specific portions of the output state. However, the oracle will fail to identify the failure if it does not check the specific part of the output state which contains the erroneous value. Therefore, the oracle should also \textit{reveal} the failure~\cite{li2016test}, as proposed in the RIPR model. This paper further elaborates this concept defining how mutation testing can be designed to validate and measure the quality of a test suite \changes{w.r.t.} a requirement, in our case taking the form of a rigorously defined STL property for a MathWorks\textsuperscript{\textregistered} Simulink model.

\section{Property-based Mutation Testing}
\label{sec3}
In this section, we present \textit{PBMT}, a mutation testing approach designed to validate test suites against programs and properties. We assume that we have a program $\mathcal{P}$ expressed in a language $\mathcal{L}$ as the software-under-test (SUT), a property $\phi$ of the SUT, a test suite $\mathcal{T}$ and a set of mutation operators $\mathcal{O}$. PBMT measures how thoroughly the test suite $\mathcal{T}$ validates $\mathcal{P}$ against the property $\phi$, studying the capability of $\mathcal{T}$ to reveal faults \textemdash of type defined in $\mathcal{O}$\textemdash that may impact $\phi$.

\begin{definition}[\textit{$\phi$-killed mutant}]\label{def:pkill}
A mutant $p$ is said to be \textit{$\phi$-killed} by a test suite $\mathcal{T} \subset \mathcal{T}_U^\mathcal{P}$ 
iff $\exists$ a test case $t \in \mathcal{T}$ such that the following conditions hold: 

\begin{enumerate}
    \item $O(t,\mathcal{P}) \models \phi$, i.e., $t$ satisfies $\phi$ when executed on the original program $\mathcal{P}$, and
    \item $O(t,p) \not\models \phi$, i.e., $t$ violates $\phi$ when executed on the mutant $p$. It follows that $t$ exercises the mutation/fault in $p$ in such a way that its effect is propagated to the output up to the violation of the property $\phi$.
\end{enumerate}

The above two conditions collectively guarantee that the execution of $p$ against $t$ yields an output strong enough to violate 
$\phi$ (i.e., $O(t,p) \not\models \phi$), while still passing in the original program (i.e., $O(t,\mathcal{P}) \models \phi$). This implies that the test is specifically good in exercising the software so that the fault, if present, is propagated to the output\changes{,} producing significant behavioral differences up to the point of violating 
$\phi$.
\end{definition}

Similar to the concept of equivalent mutants in regular \changes{MT}, we introduce a refined version of equivalent mutants which we call: \textit{$\phi$-trivially different mutants}. The intuition is that in this context, a mutant is irrelevant not only if it is equivalent (i.e., it shows no behavioral differences \changes{w.r.t.} the original program), but also if the introduced behavioral differences are not relevant \changes{w.r.t.} the property $\phi$, that is, no test case $t \in \mathcal{T}_U^\mathcal{P}$ can distinguish between $p$ and $\mathcal{P}$.

\begin{definition}[\textit{$\phi$-trivially different mutant}]
A mutant $p$ is \textit{$\phi$-trivially different} from $\mathcal{P}$ 
iff $\nexists t \in \mathcal{T}_U^\mathcal{P} : O(t,\mathcal{P})\models \phi \land O(t,p) \not\models \phi$.
\end{definition}

The set of the \textit{$\phi$-trivially different} mutants include equivalent mutants. The identification of \textit{$\phi$-trivially different mutants} is undecidable. 

\begin{definition}[\textit{$\phi$-adequate test suite}]
A test suite $\mathcal{T}$ is \textit{$\phi$-adequate} w.r.t. a set of mutation operators $\mathcal{O}$ if it kills all the non $\phi$-trivially different mutants that can be generated by $\mathcal{O}$.
\end{definition}


\begin{definition}[\textit{Mutation score}]\label{def:MS}
If $KD_{\phi}$ denotes the $\phi$-killed mutants and $NTD_{\phi}$ denotes the non $\phi$-trivially different mutants, the mutation score assigned with a test suite $\mathcal{T}$ for a program $\mathcal{P}$ and a set of mutation operators $\mathcal{O}$ is


\begin{equation}
    \mathcal{MS_{\phi}} = \frac{|KD_{\phi}|}{|NTD_{\phi}|}
    \label{eq1}
\end{equation}%
\end{definition}

The objective of implementing test suites that are adequate according to PBMT results in the \textit{Mutant killing problem}. That is, given a program $\mathcal{P}$, a mutant of $\mathcal{P}$ denoted by $p$ and a property $\phi$, the \textbf{mutant killing problem} is the problem of finding a test case $t$ such that $O(t,\mathcal{P})\models \phi$, and $O(t,p) \not\models \phi$.

PBMT is usually more challenging than regular MT since:


\begin{itemize}
    \item \textit{Higher risks of introducing $\phi$-trivially different mutants}: PBMT can potentially generate more irrelevant mutations than mutation testing since, in addition to equivalent mutants, there might be mutants that are not equivalent but introduce irrelevant differences w.r.t. a property $\phi$.
    \item \textit{Harder to kill mutants}: The faults must be exercised in such a way that it does not only propagate to the output but also leads to the violation of $\phi$. 
\end{itemize}


\section{Mutation Testing of Simulink CPS Programs}\label{sec4}
We instantiate PBMT in the context of safety-critical CPS Simulink (data-flow) models where the system safety properties are expressed using STL. While extensive details of Simulink models~\cite{chowdhury2018automatically,chowdhury2018curated,rajhans2018graphical} and STL~\cite{maler2004monitoring,maler2013monitoring,donze2010robust} are available elsewhere, we introduce below the key concepts to make the paper self-contained. We conclude by presenting a novel technique to automatically determine the mutants that could be $\phi$-killed by test suites.

\subsection{Simulink models}
\begin{figure}[htbp]
\centerline{\includegraphics[width=0.5\linewidth]{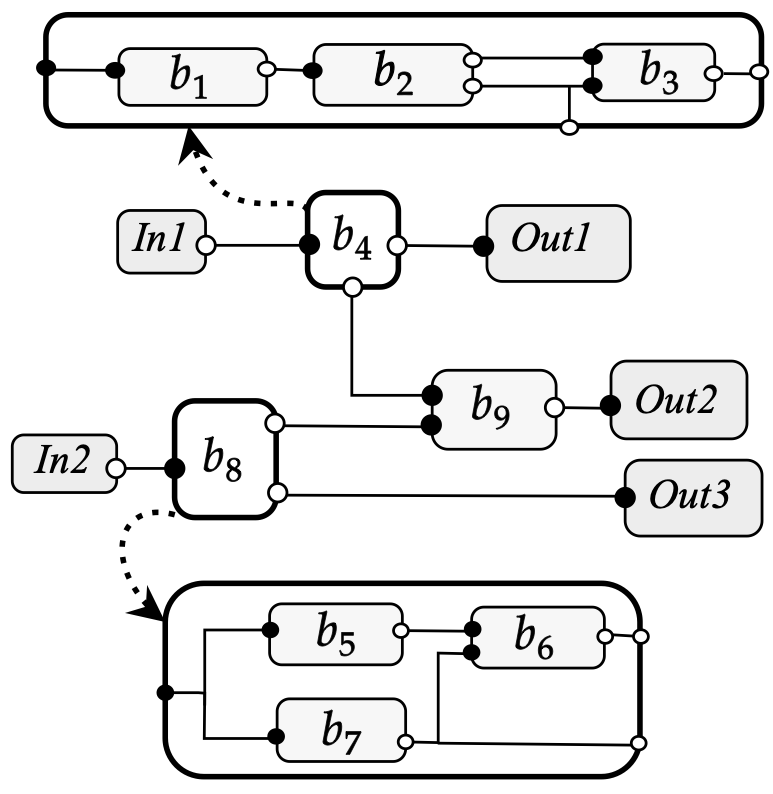}}
\caption{A Simulink model with hierarchical blocks (\changes{$b_4, b_8$}) and atomic blocks (remaining), input ports (black nodes), output ports (white nodes), inputs (\textit{In1} and \textit{In2}) and outputs (\textit{Out1}, \textit{Out2} and \textit{Out3}).}
\label{fig:simbd}
\end{figure}

The MathWorks\textsuperscript{\textregistered} Simulink environment is widely used for  CPS model-based development~\cite{Simulink,MBSEng}. Simulink allows non-software engineers to design complex systems, compile them to low-level code, and simulate the designed models to observe their behavior against some test inputs. In general, a Simulink model 
is the block diagram representation of a system using blocks and lines (aka \textit{connections}) as in Fig.~\ref{fig:simbd}. 
A block receives data via its \textit{input ports} and performs a defined operation on its input data depending on its functionality. After processing the input data, a block transmits the output data via its \textit{output ports}, along (directed) lines. Each line in the model can be uniquely identified using (1) the source block and its associated output port, and (2) the target block and its associated input port. The model receives its inputs from a set of input blocks and emits the output through a set of output blocks. Usually, a block can be either \textit{atomic} (i.e., it does not include any other block within it) or \textit{hierarchical} (i.e., it includes other blocks within it). 

When creating a model, a tester can either use standard blocks from built-in \textit{libraries} or create new custom blocks from scratch. After designing the model, a tester \textit{compiles} and \textit{simulates} the model \changes{using a suitable} 
\textit{solver} and \textit{simulation mode}. Simulink allows to execute the model using user-specified \textit{sample times} (either fixed-length or variable-length). 

A Simulink model $\mathcal{M}$ when simulated against a test case $t$ yields the model simulation output as the set of traces of all input-internal-output signals. We denote the model simulation output with $O(t,\mathcal{M})$. A Simulink model can have multiple outputs (such as Fig.~\ref{fig:simbd}'s \textit{Out1}, \textit{Out2} and \textit{Out3}). 

\subsection{Signal Temporal Logic (STL)} 
In recent years, for 
the verification of safety-critical CPS, researchers have used temporal logic formalisms to express safety properties. Signal Temporal Logic (STL)~\cite{maler2013monitoring} is a well-known specification formalism used to express temporal properties of dense-time real-valued behaviors of hybrid (i.e., both continuous and discrete dynamic) systems, including safety-critical CPS. The syntax of STL is formally defined as follows:
\begin{displaymath}
\Phi := f(\mathbf{x}(j)) > 0 \;|\; \lnot \Phi \;|\; \Phi_1 \land \Phi_2 \;|\; \square_I \Phi \;|\; \lozenge_I \Phi \;|\; \Phi_1  \mathcal{U}_I \Phi_2
\end{displaymath}
Here, the formula of the form $f(\mathbf{x}(j)) > 0$ represents a \textit{signal predicate}, where $\mathbf{x}(j)$ is the value of a signal $\mathbf{x}$ at time instant $j$, and $f$ is a function from signal domain $\mathcal{D}$ to $\mathbb{R}$. $I \subseteq \mathbb{R}_{\geq 0}$ is an arbitrary time-interval. The propositional logic operators $\lnot$ and $\land$ follow the obvious logical semantics, i.e., $\lnot$ indicates logical negation and $\land$ indicates logical conjunction. Other temporal operators are as follows:
\begin{itemize}
    \item ${\square_I \Phi}$  (\textit{always} operator) indicates that $\Phi$ must be true for all samples in $I$.
    \item $\lozenge_I \Phi$ (\textit{eventually} operator) indicates that $\Phi$ must be true \textit{at least once} for samples in $I$.
    \item $\Phi_1  \mathcal{U}_I \Phi_2$ means that $\Phi_1$ must be true in $I$ until $\Phi_2$ becomes true.  $\mathcal{U}_I$ refers to as \textit{until} operator. 
\end{itemize}

The Boolean satisfaction semantics aka \textit{qualitative semantics} of STL offers a boolean witness of the property $\Phi$. The Boolean satisfaction of the signal predicate is simply $\top$ if it is satisfied; otherwise $\bot$. We use the operators $\mathcal{U}$, $\lozenge$, and $\square$ to denote $\mathcal{U}_I$, $\lozenge_I$, and $\square_I$ with $I=[0,\infty)$.

Besides the qualitative semantics, STL also offers quantitative semantics~\cite{donze2010robust} that allows to compute the degree of satisfaction of $\Phi$ by the traces generated by a system after executing it against a test input. The degree of satisfaction of $\Phi$ for a trace $q$ is measured using a \textit{robust satisfaction function}  $\rho(q,\Phi)$ that computes a real value that indicates the distance of the trace $q$ from satisfying ($\models_s$) 
the property $\Phi$.  Formally, $\rho(q,\Phi) > 0 \Rightarrow q \models_s \Phi$, and $\rho(q,\Phi) < 0 \Rightarrow q \not\models_s \Phi$.

\subsection{Mutations in Simulink}
\begin{figure}[htbp]
\centerline{\includegraphics[width=1\linewidth]{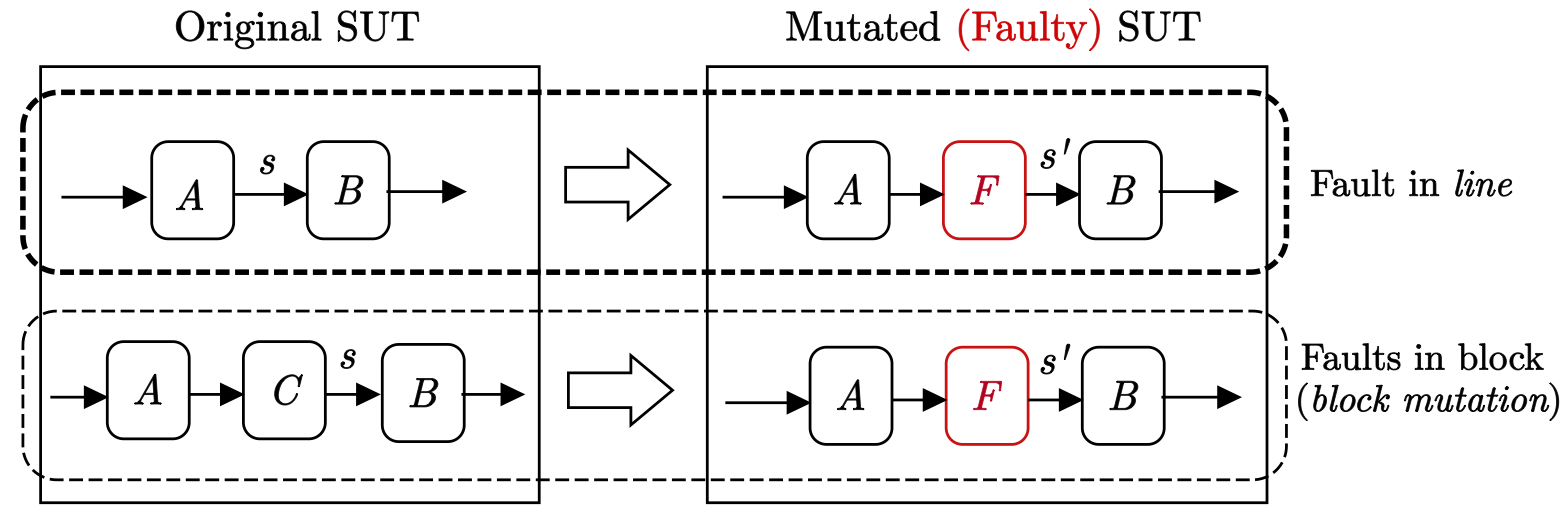}}
\caption{Mutations in a SUT (the seeded fault blocks $F$ are highlighted in red). $A$, $B$ and $C$ are blocks of original SUT. Internal signals $s$ and $s'$ provide knowledge of the fault location.}
\label{fig:F}
\end{figure}

From a conceptual perspective, mutations 
are simply modifications to the behavior of the Simulink model. Usually, alterations can be made in a Simulink model in two ways: 
\begin{enumerate}[leftmargin=*]
    \item \emph{Line mutations}: changing the behavior of the signals that propagate through lines from one block to another block (see `Fault in \textit{line}' in Fig.~\ref{fig:F}), or
    \item \emph{Block mutations}: changing the behavior of a block (see `Fault in block' in Fig.~\ref{fig:F}), for instance, by making changes in its functionality.
\end{enumerate}

\vspace{-2mm}

\subsection{Robustness Measure}
The notion of \textit{robustness} function $\rho$ becomes 
useful when we need to search for a test $t$ that passes the execution of the model $\mathcal{M}$ w.r.t. an STL requirement $\phi$.
We use the following notations~\cite{donze2010robust}: 
 \begin{enumerate}
     \item $\rho (O(t,\mathcal{M}),\phi) < \epsilon \Rightarrow O(t,\mathcal{M}) \not\models \phi$, i.e., $t$ fails on $\mathcal{M}$ with respect to the specification $\phi$
     \item $\rho (O(t,\mathcal{M}),\phi) > \epsilon \Rightarrow O(t,\mathcal{M}) \models \phi$, i.e., $t$ passes on $\mathcal{M}$ with respect to the specification $\phi$.
 \end{enumerate}

 Here, the parameter $\epsilon$ represents the degree of violation of the property as assessed by the robustness function $\rho$. The standard choice is $\epsilon = 0$ which implies that the identification of passing or failing test case i.e., satisfaction or violation 
 is based on even a small (non-zero) deviation in the observed behavior of $\mathcal{M}$ from the expected behavior w.r.t. $\phi$.

\subsection{Search-based generation of mutation adequate test cases}

A key challenge in mutation testing, including PBMT, is accurately computing the mutation score, due to the undecidable problem of identifying the equivalent mutants. In PBMT, this problem is even harder due to the need of identifying the $\phi$-trivially different mutants, which include but are not limited to the equivalent mutants. To address this challenge, we defined a search-based test generation strategy that exploits the knowledge of the mutants and their locations to generate targeted executions that demonstrate if a mutant can be $\phi$-killed. Although nothing could be said about the mutants not killed according to this procedure, the experimental results show that assuming this procedure can identify every $\phi$-killable mutant may give an accurate approximation of the mutation score. 

Note that the proposed test strategy cannot be used to generate tests in a real situation, since it exploits the knowledge of the fault location that is normally unknown when a software is tested. However, the proposed test generation strategy is useful in the context of PBMT to collect accurate empirical data.

In particular, we formulate the `Property-based test search problem', an optimization problem of finding a $\phi$-adequate test case as:

\begin{tcolorbox}[width=\linewidth,title=Property-based test search problem,outer arc=0mm]    
    \textbf{\textsc{Input}}: a Simulink model $\mathcal{M}$, a first-order mutant $\mathcal{M}'$ (with signal $s$ changed into signal $s'$ or a block $b$ with output $s$ changed into a block $b'$ with output $s'$),  and a property $\phi$. \\
     \textbf{\textsc{Problem}}: Find $t$ s.t. $\rho(O(t,\mathcal{M}),\phi) > 0$, $\rho(O(t,\mathcal{M'}),\phi) < 0$ and $D(s, s')$ is maximum.
\end{tcolorbox}

The proposed `Property-based test search problem' combines three key \changes{features}, two deriving from the definition of $\phi$-killed mutant and one guiding the search toward the mutant, and toward producing an execution that exploits the mutant to significantly alter the state of the system:
\begin{itemize}
\item $\rho(O(t,\mathcal{M}),\phi) > 0$ requires finding a test that passes on the original program,
\item $\rho(O(t,\mathcal{M'}),\phi) < 0$ requires finding a test that violates $\phi$ in the modified program, and
\item $D(s, s')$ \emph{is maximum} requires the mutation to impact on the internal signal as much as possible.
\end{itemize}


We choose the Euclidean distance (aka $L^2$ norm) as the metric to compute the distance between $s$ and $s'$. Since CPS models involve continuous real-valued variables, Euclidean distance, a prominent metric for real vector spaces, is a good candidate for computing the distance. More rigorously, given two finite-length signals $s=(s_1,\cdots,s_k)$ and $s'=(s_1',\cdots,s_k')$, each with $k$ samples, the Euclidean distance between $s$ and $s'$ is mathematically expressed as:
\begin{displaymath}
  D(s,s')= ||s-s'||_2= \sqrt{\sum_{i=1}^{k} (s_i-s_i')^2}
\end{displaymath}

\changes{The optimization task is to maximize $D(s, s')$ subject to the constraints $\rho(O(t,\mathcal{M}),\phi) > 0$ and $\rho(O(t,\mathcal{M'}),\phi) < 0$.} To solve the formulated test search problem, we exploit BCA~\cite{yadav2021blood}, a recently developed global optimizer as outlined in Algorithm~\ref{algoST}. \changes{We chose BCA over other available optimizers on account of its superior convergence and speed.} While being a global search with BCA in essence, Algorithm~\ref{algoST} introduces two differences \changes{w.r.t.} standard BCA: (1) The initial population (Line 2) is a set of test cases randomly generated in their valid numerical input domain. (2) Fitness (Line 3) corresponds to the value of the test objective function for the given population of test cases. \changes{The test objective function is obtained by converting the constrained optimization problem into an unconstrained problem using the scalar penalty constraint handling method~\cite{coello2002theoretical}.} The algorithm updates the test cases (Line 6-8) and finds the best solution for the new population depending on their fitness values (Lines 9-10). The candidate fittest amongst all others in the population is accepted as the new global best solution 
(Lines 11-14). The algorithm returns the best solution if all the constraints are satisfied. Algorithm~\ref{algoST} terminates (loop at Line 5) if either a test case satisfying the optimization constraints is found, or the budget is exhausted (time budget or the maximum number of iterations).  

\begin{algorithm}
    \SetKwInOut{KwIn}{Input}
    \SetKwInOut{KwOut}{Output}
    \KwIn{$\mathcal{M}$ : A Simulink model.\\ $\mathcal{M'}$ : A mutant of $\mathcal{M}$. \\
    $\phi$ : An STL specification.}
    \KwOut{$t_{best}$ : A test case that $\phi$-kills $\mathcal{M'}$.}
    
    Initialize optimizer parameters\\
    $\mathsf{IP} \leftarrow \textsc{GenerateInitialPopulation}()$\
    
    $\mathsf{FP} \leftarrow Fitness(\mathsf{IP}, \mathcal{M}, \mathcal{M'}, \phi)$ \\ 
    
    $t_{best}, F_{best} \leftarrow BestFound(\mathsf{FP})$ \\ 
    
    \While {$\mathsf{TimeOut}()$}{
        \For{each candidate $k \in \mathsf{IP}$}{
            $k_{new} \leftarrow Update(k)$
        }
        $\mathsf{FP} \leftarrow Fitness(\mathsf{IP},\mathcal{M}, \mathcal{M'},\phi)$\\
        $t_{new}, F \leftarrow BestFound(\mathsf{FP})$ \\
        \If{$F > F_{best}$}{
            $F_{best} \leftarrow F$ \tcp*{update best fitness}
            $t_{best} \leftarrow t_{new}$
            \tcp*{update best test}
        }
    }
    \KwRet{$t_{best}$}
    \caption{Search-based test generation.}
    \label{algoST}
\end{algorithm}

For each mutant, we solve the formulated `Property-based test search problem' to find a test case that $\phi$-kills it. The resulting test suite is a \emph{fault-directed test suite} that is likely to reveal all the \emph{non $\phi$-trivially different} mutants.

\subsection{Test suite reduction}
To maintain a small and practical fault-directed test suite, we reduce its size automatically. We consider a test case $t_r$ $\phi$-\emph{redundant} w.r.t. a fault-directed test suite $\mathcal{T}$ if the set of $\phi$-\textit{killed} mutants by $\mathcal{T}$ remains unchanged after the inclusion of $t_r$ in $\mathcal{T}$, i.e., \changes{$|KD_{\phi}|_{\mathcal{T}} = |KD_{\phi}|_{\mathcal{T}\,\cup\,t_r}$.} 

A $\phi$-\textit{non-redundant} test suite does not contain $\phi$-redundant test cases. Usually, a test suite can contain redundant test cases while retaining the same \textit{testing power} in the sense that they are capable of killing the same mutants w.r.t. $\phi$. In other words, a single test case can cover more than one mutation. 


In our experiments, we use the greedy algorithm similar to the one proposed in~\cite{polo2012reduction} for test suite reduction. In the worst-case scenario, $p$ test cases are required to cover all $p$ non $\phi$-trivially different mutations. In practice, fewer tests are usually necessary.

\section{Evaluation}\label{sec5}
Our evaluation aims to study Property-Based Mutation Testing (PBMT) for testing CPS Simulink models against STL properties, also w.r.t. regular Mutation Testing (MT).  


\subsection{Research Questions}
Our experiments address the following research questions:

\textbf{RQ1.} \textit{Does PBMT measure the adequacy of a test suite better than MT when a safety property is targeted?} To answer this research question, we assess the adequacy of multiple test suites using both PBMT and MT, and discuss how the resulting scores reflect the intrinsic capability of the test cases to exercise the software based on the target property.



\textbf{RQ2.} \textit{Are mutation operators equally contributing in PBMT?} To answer this research question, we study the impact of different mutation operators on the mutation score, aiming at discovering operators that tend to generate mutants that are either trivial or particularly hard to detect.

\subsection{Experimental Setup}
We performed our experiments on a MacBook Pro with Apple M1 chip, 16 GB RAM, macOS Monterey with MATLAB{\texttrademark} R2018b. For our evaluation, we developed a prototype implementation of both PBMT and MT with CPS Simulink models in MATLAB. We used the RTAMT library~\cite{DBLP:conf/atva/Nickovic020} for offline evaluation of STL properties. 

We limit the scope of the evaluation to FOMs\changes{. }
Moreover, we use a fixed-length sampling when running Simulink models with faults active from the beginning to the end of the simulation. 
In the following, we describe our experimental subjects, mutants and test suites.

\subsubsection{Experimental subjects}
We evaluate PBMT on Simulink models of two industrial benchmarks across the safety-critical domain, each one publicly available in the  Simulink/Stateflow online documentation of MathWorks\textsuperscript{\textregistered}~\cite{ATCS,AECS}: \textit{ATCS}, an Automatic Transmission Controller System, and  \textit{AECS}, an Aircraft Elevator Control System. 

ATCS is a typical automotive drivetrain with the two inputs \emph{throttle} and \emph{brake} governing the vehicle speed $v$ (mph) and the engine speed $\omega$ (RPM). Both user inputs are in the range [0, 100] for all time instants. As one of the safety properties, ATCS requires that $v$ and $\omega$ must always remain below their thresholds $\bar{v}$ and $\bar{\omega}$, respectively. This is represented in STL in Table~\ref{tab:IND} where $\bar{v} = 120$ mph and $\bar{\omega} = 4500$ RPM.

AECS from the avionics-aerospace domain controls the positions of the left and right elevators of an aircraft using the pilot command. In general, the elevator position should maintain a constant value if the aircraft is flying at the desired level. Among the safety requirements, the AECS requires that whenever the Pilot Command \textit{cmd} goes beyond a threshold \textit{m}, the measured elevator position \textit{pos} must stabilize (should not exceed \textit{cmd} by more than \textit{n} units) within $T + a$ time units. This is formally expressed with the STL specification in Table~\ref{tab:IND} where $m=0.09$, $T=2$, $a=1$ and $n=0.02$. 

\begin{table*}
  \caption{Details of Simulink models of our case studies.}
  \label{tab:IND}
  \centering
  \begin{tabular}{llllllll}
    \toprule
    Model & Ref. & \#$Blocks$ & \#$Lines$ & $\phi$ (STL specification) & $q_T$ & Sample time & \#$Samples$\\
    \midrule
    ATCS & \cite{donze2010breach} & 65 & 92 & $ \square((v \leq \bar{v}) \land (\omega \leq \bar{\omega}))$ & 30 & 0.04 & 751 \\
    AECS & \cite{bartocci2021cpsdebug} & 825 & 577 & $\square(\uparrow (cmd \geq m) \rightarrow \lozenge_{[0,T]} \square_{[0,a]} (|cmd - pos| \leq n))$ & 10 & 0.01 & 1001\\
    \bottomrule
 \end{tabular}
\end{table*}

\subsubsection{Fault seeding and mutant generation}
For each experimental subject, we generated mutants using the FIM prototype tool~\cite{bartocci2022fim} that supports the following mutation operators for Simulink models: 
    \texttt{Negate}, \texttt{Stuck-at}, \texttt{Absolute}, \texttt{Noise}, \texttt{Bias/Offset}, \texttt{Time Delay}, \texttt{Package Drop}, 
\texttt{ROR} (Relational Operator Replacement), \texttt{LOR} (Logical Operator Replacement), \texttt{S2P} (Sum to Product mutation), \texttt{P2S} (Product to Sum mutation) and \texttt{ASR} (Arithmetic Sign Replacement). \changes{The detailed description of these operators can be found in~\cite{bartocci2022fim}.}

Since FIM does not support the injection of faults in look-up tables (LUTs), we extended the tool implementing two additional operators: (1) Stuck-at 0 fault in any one entry, 
and (2) swapped entries (from two randomly chosen neighbors).

Table~\ref{tabfaultpattern} reports, for each subject, the number of mutants generated for the specific mutation operator. Table~\ref{tabfault} indicates the total number of mutants generated for every subject and their generation time. Mutant generation is fast: 
On an average (across ATCS and AECS), the generation of a mutant takes 1.74 seconds. 

\begin{table}[htbp]
\begin{center}
\caption{Number of mutants of our experimental subjects.}
\label{tabfaultpattern}
\begin{tabular}{lcc}
\toprule
\multirow{2}{*}{\textbf{Type}} & \multicolumn{2}{c}{\textbf{\# Mutants}}\\
\cmidrule(lr){2-3}
& {ATCS} & {AECS}\\
\cmidrule(lr){1-1}\cmidrule(lr){2-2}\cmidrule(lr){3-3}
\texttt{Noise} & 13 & 17\\
\texttt{Bias/Offset} & 13 & 17\\
\texttt{Negate} & 13 & 17\\
\texttt{Absolute} & 13 & 17\\
\texttt{ROR} & 0 & 10\\
\texttt{S2P} & 1 & 3\\
\texttt{P2S} & 2 & 6\\
\texttt{ASR} & 3 & 8\\
\texttt{LUT} & 2 & 5\\
\bottomrule
\end{tabular}
\end{center}
\end{table}

\vspace{-2mm}

\begin{table}[htbp]
\begin{center}
\caption{Information of generated mutants.}
\label{tabfault}
\begin{tabular}{ccc}
\toprule
\textbf{Subject} & \textbf{Mutants generated}	& \textbf{Mutant generation time (seconds)}\\
\cmidrule(lr){1-1}\cmidrule(lr){2-2}\cmidrule(lr){3-3}
ATCS & 60 & 68.76\\
AECS & 100 & 261.64\\
\bottomrule
\end{tabular}
\end{center}
\end{table}

\subsubsection{Test Suite}
To compare PBMT to MT, we assess test suites generated according to two different strategies: Adaptive Random Testing (ART)~\cite{liu2019effective} and Falsification Testing (FT)~\cite{DBLP:conf/cav/AdimoolamDDKJ17,DBLP:conf/cav/AkazakiH15}. ART is a baseline strategy that generates evenly distributed test cases (within valid input ranges), thereby ensuring adequate diversity in the test inputs. On the other hand, FT generates counterexamples i.e., test cases that violate a property for a given model~\cite{DBLP:conf/cav/ZhangHA19,DBLP:conf/cav/ZhangLAMHZ21}. Note that ART and FT work in radically complementary ways. ART quickly generates many test inputs, considering diversity, but ignoring the property under test. On the contrary, FT specifically targets the generation of a test that violates the property under test. In particular, for each mutant $\mathcal{M'}$, FT attempts to generate a test case $t$ such that $O(t,\mathcal{M'}) \not\models \phi$. The hypothesis is that ART could obtain higher $\mathcal{MS}$, but smaller $\mathcal{MS}_\phi$ since the generated tests do not depend on $\phi$. On the contrary, FT should kill fewer mutants in general, but more mutants relevant to $\phi$, and thus obtain higher $\mathcal{MS}_\phi$.

In our evaluation, we generated 30 and 50 test cases with ART for ATCS and AECS, respectively. FT generates a property-violating test per mutant, if successful. 

For collecting data to address our research questions, we have executed all the test cases in the test suite for every subject and every generated mutant. To perform our experiments, we executed multiple simulations in parallel using the Parallel Computing Toolbox™ in the MATLAB/Simulink\textsuperscript{\textregistered} environment. Table~\ref{tabscale} provides, for each subject, the total number of test cases executed (including both test suites) and the total execution time. 

\begin{table}[htbp]
\begin{center}
\caption{Scale of Experiments.}
\label{tabscale}
\begin{tabular}{ccc}
\toprule
\textbf{Subject} & \textbf{Total test cases executed}	& \textbf{Total execution time (seconds)}\\
\cmidrule(lr){1-1}\cmidrule(lr){2-2}\cmidrule(lr){3-3}
ATCS & 90 & 2,490\\
AECS & 150 & 25,912\\
\bottomrule
\end{tabular}
\end{center}
\end{table}



\subsection{Results}

\begin{table*}[htbp]
\begin{center}
\caption{Results of Mutation Testing.}
\label{tabCMAPBMT}
\begin{tabular}{llcccc}
\toprule
\multirow{2}{*}{\textbf{Approach}} & & \multicolumn{2}{c}{\textbf{ATCS}} & \multicolumn{2}{c}{\textbf{AECS}} \\
\cmidrule(lr){3-4}
\cmidrule(l){5-6}
& & $\mathcal{T}_{ART}$ & $\mathcal{T}_{FT}$ & $\mathcal{T}_{ART}$ & $\mathcal{T}_{FT}$\\
\cmidrule(lr){1-2}
\cmidrule(lr){3-4}
\cmidrule(l){5-6}
\multirow{4}{*}{MT} & \# Mutants & 60 & 60 & 100 & 100\\
& \# \textit{Killable} mutants & 47 & 47 & 83 & 83\\
& \# \textit{Killed} mutants & 47 & 46 & 74 & 70\\
& Mutation Score $\mathcal{MS}$ (in \%) & 100\% & 97.87\% & 89.15\% & 84.33\%\\

\midrule

\multirow{4}{*}{PBMT} 
& \# Mutants & 60 & 60 & 100 & 100\\
& \# \textit{$\phi$-killable} mutants & 47 & 47 & 83 & 83\\
& \# \textit{$\phi$-killed} mutants & 25 & 27 & 39 & 35\\
& $\mathcal{MS_\phi}$ (in \%) & 53.19\% & 57.44\% & 46.98\% & 42.16\%\\
\bottomrule
\end{tabular}
\end{center}
\end{table*}
\textbf{RQ1} studies the extent to which PBMT-based testing can better capture the thoroughness of a test suite \changes{w.r.t.} a safety property that the software-under-test must fulfil. To this end, we apply both MT and PBMT to our experimental subjects and compute the mutation scores $\mathcal{MS}$ and $\mathcal{MS_\phi}$. Note that we use exactly the same mutants to compute both scores. Table~\ref{tabCMAPBMT} reports the results. 

We report the results for regular mutation testing (MT) and Property-Based Mutation Testing (PBMT) in two different rows, while columns ATCS and AECS correspond to the two subject systems. For each subject system, we indicate the scores achieved by the test suites generated with Adaptive Random Testing ($\mathcal{T}_{ART}$) and Falsification Testing ($\mathcal{T}_{FT}$). In details, we report the number of mutants that have been generated, the number of killable and $\phi$-killable mutants, the number of mutants killed by each test suite according to MT and PBMT, and finally the mutation scores $\mathcal{MS}$ and $\mathcal{MS}_\phi$.

To identify the killable mutants, we had to identify the equivalent ones. To this end, we inspected the non-killed mutants to determine if a mutation generated a variant that cannot be distinguished from the original program. We could identify every equivalent mutant with high-confidence. In fact, the 13 equivalent mutants in the ATCS model all belong to the \texttt{Absolute} fault type injected in the `Transmission' component and all try to change into positive values some signals that could not be negative. The exact same situation happened for the 17 equivalent mutants found in the AECS model. To determine the $\phi$-killable mutants, we used the Search-based test generation (SBTG) technique presented in Algorithm~\ref{algoST}. \changes{Note that the SBTG strategy is more computationally expensive than ART and FT due to the optimization constraints.} Our procedure automatically identified every $\phi$-killable mutant with thirty independent runs of our search algorithm and a maximum number of iterations (set to 1000) as the stopping criterion. The remaining $\phi$-trivially different mutants are all equivalent mutants that cannot be killed. This result provides confidence on the capability of our approach to support fully automated experiments with Simulink models by assuming that the mutants not killed with our strategy are $\phi$-trivially different mutants that do not need to be killed, and thus can be excluded from the computation of $\mathcal{MS}_\phi$.

By comparing the results obtained for MT to the results obtained with PBMT, we can notice the mutation score obtained with MT is significantly higher than the mutation score obtained with PBMT. In fact, the value of $\mathcal{MS}$ ranges between $84.33\%$ and $100\%$ for the four test suites and the two subject systems. On the other hand, the value of $\mathcal{MS}_\phi$ ranges between $42.16\%$ and $57.44\%$. 
This is also due to the intrinsic nature of both Simulink models and data-flow computations, where it is generally easy to activate every component (i.e., to generate a sequence of inputs that exercise every element in a program), but it is definitely harder to activate these components while  guaranteeing they contribute to the computation propagating the fault to the output, finally causing observable issues. That is, it is relatively easy to \emph{reach} faults, but it is still hard to meaningfully \emph{propagate} and \emph{detect} faults. This result is confirmed across the test suites generated with two alternative strategies.

These results demonstrate that MT may mislead testers when there are important properties to be validated. \changes{For instance, referring to 
Fig.~\ref{fig:1} (top), the test case can \textit{kill} the mutant but cannot $\phi$\textit{-kill} it.} In fact, the test suites generated with ART and FT achieve high mutation score ($\mathcal{MS}$), possibly inducing testers to believe the test suites are thoroughly exercising software. On the contrary, it turns out that the test cases are not good enough to guarantee that even the simple faults (e.g., like the ones we injected) that may affect the property are actually detected.

It is also interesting that FT, which targets the falsification of the property, in comparison to ART, which addresses diversity neglecting the existence of the property, does not kill more mutants. Combined with the evidence that almost half of the killable mutants have not been $\phi$-killed, this suggests that more research is needed to exercise software thoroughly \changes{w.r.t.} a target property, at least for Simulink programs.

We finally checked for the capability of the generated tests to kill and $\phi$-kill mutants. Interestingly, there is often high redundancy across tests, that is, each test can kill many mutants. For instance, all the mutants that have been killed with ART could be killed by a single test. This reinforces the idea that there are some surface faults that are easy to reveal, but at the same time there are other faults that, even if simple in structure, require more sophisticated tests to be revealed.

On the other hand, we found that
four test cases, derived with our SBTG technique are needed to reveal all 47 $\phi$-killable mutations of ATCS. Likewise, all 83 $\phi$-killable mutations of AECS could be revealed with 12 test cases. This suggests that compact but effective test suites could be designed to reveal faults according to PBMT. Yet, PBMT requires a higher number of tests 
than regular MT 
to $\phi$-kill and kill mutants, respectively.

\textbf{RQ2} assesses the contribution of individual mutation operators in PBMT. The goal is to identify the operators that tend to generate easy-to-kill mutants (\emph{simple mutants}), which do not contribute much to measuring the adequacy of a test suite, and the operators that tend to generate hard-to-kill mutants (\emph{stubborn mutants}), which can contribute more in measuring the thoroughness of a test suite. 

Table~\ref{tabPBMT_OP} reports the following results for each mutation operator: (1) the number of mutants generated, (2) the number (and percentage) of $\phi$-trivially different mutants, (3) the number (and percentage) of $NTD_\phi$ (i.e., non $\phi$-trivially different mutants), (4) mutation score achieved by ART, (5) mutation score achieved by FT, and (6) number (and percentage) of $NTD_\phi$ mutants \textit{not killed} by any test generation technique (neither ART nor FT). Note that Table~\ref{tabPBMT_OP} reports the combined results for our two experimental subjects (ATCS and AECS).

\begin{table*}[htbp]
\begin{center}
\caption{Summary of results of PBMT for individual operators.}
\label{tabPBMT_OP}
\begin{adjustbox}{width=1\textwidth}
\small
\begin{tabular}{llllllllll}
\toprule
& \texttt{Noise} & \texttt{Negate} & \texttt{Bias} & \texttt{Absolute} & \texttt{ROR} & \texttt{S2P} & \texttt{P2S} & \texttt{ASR} & \texttt{LUT}\\
\midrule
\# Mutants generated & 30 & 30 & 30 & 30 & 10 & 4 & 8 & 11 & 7\\
\# (\%) of $\phi$-trivially different mutants & 0 (0\%) & 0 (0\%) & 0 (0\%) & 30 (100\%) & 0 (0\%) & 0 (0\%) & 0 (0\%) & 0 (0\%) & 0 (0\%)\\
\# (\%) $NTD_\phi$ 
& 30 (100\%) & 30 (100\%) & 30 (100\%) & 0 (0\%) & 10 (100\%) & 4 (100\%) & 8 (100\%) & 11 (100\%) & 7 (100\%)\\
$\mathcal{MS_\phi}_{ART}$ (in \%) & 66.67\% & 43.33\% & 46.66\% & 0\% & 0\% & 25\% & 62.5\% & 45.45\% & 85.71\%\\
$\mathcal{MS_\phi}_{FT}$ (in \%) & 70\% & 43.33\% & 50\% & 0\% & 0\% & 25\% & 62.5\% & 45.45\% & 28.57\%\\
\# (\%) $NTD_\phi$ \textit{not killed} by ART+FT & 9 (30\%) & 17 (56.66\%) & 15 (50\%) & 0 (0\%) & 10 (100\%) & 3 (75\%) & 3 (37.5\%) & 6 (54.54\%) & 1 (14.28\%)\\
\bottomrule
\end{tabular}
\end{adjustbox}
\end{center}
\end{table*}

At least half of the mutations generated by the \texttt{Negate}, \texttt{ROR}, \texttt{S2P} and \texttt{ASR} operators have been killed neither by ART nor by FT. This may suggest that these operators might be more useful than others for PBMT because they tend to generate faults that are not easy to propagate to the output. 


For instance, all the mutants of AECS with the \texttt{Negate} operator were generated by alterations in the \textit{Right Outer Hydraulic Actuator} component. The available test cases can easily infect the execution (e.g., they change the output of the `Line resistance' block), but fail to propagate the infection due to the presence of an intermediate signal  (e.g., 
`Piston Force') that masks changes if differences are not large enough.

None of the mutants generated by \texttt{ROR} has been detected by $\mathcal{T}_{ART}$ and $\mathcal{T}_{FT}$. In particular, we observe that for all available test cases $t \in \mathcal{T}_{ART} \cup \mathcal{T}_{FT}$, with the execution of the \texttt{ROR} mutations, the robustness value evaluated for the STL property for every mutant is the same as that obtained for the original model. However, there exist test cases that produce visible differences in the outputs and \changes{$\phi$-}kill the mutants as demonstrated by the tests obtained with our SBTG technique.

Mutations generated by \texttt{S2P} have been also hard to \changes{$\phi$-}kill. 
Besides, some mutations with \texttt{ASR} operator could not be detected by test cases in $\mathcal{T}_{ART}$ and $\mathcal{T}_{FT}$. Though these mutants alter the internal signal, the data-flow computations and propagation of signals do not affect the property. For instance, the \texttt{ASR} mutation in the `Hydraulic Actuator' component of \textit{Right Inner Hydraulic Actuator} unit of AECS ($-+$ replaced by $+-$) creates significant variations in the local signal but is not strong enough to $\phi$-kill the mutant.


On the other hand, two operators have not been particularly useful. The \texttt{Absolute} operator only generated equivalent mutants. This suggests that this operator must be used carefully, only with systems known to process negative values, and possibly controlling the locations where the fault is injected. This case is quite infrequent in CPS. In fact, we have not observed any useful mutation in our two subjects. All the mutations generated by \texttt{LUT} were easy to \changes{$\phi$-}kill, with only one exception, which generates values hard to propagate to the output (but still feasible to propagate as demonstrated by the test suites generated with our SBTG approach). Although this operator is the only one targeting look-up tables, testers might consider skipping it when there are strong time constraints on the testing process.

\section{Threats to validity}\label{sec7}
We now discuss the threats to validity centered around the following 
perspectives of validity and threats:

\textbf{External validity.} The main threat to external validity concerns with the generalization of our results. Indeed, the reported evidence may not generalize to every software system. In fact, we experimented in the domain of data-flow oriented computations (i.e., Simulink models), and our observations may not hold in other contexts (e.g., object-oriented programs). However, results are already quite clear and explainable in the domain of safety-critical CPS Simulink programs, where testing software against safety properties is particularly relevant. Moreover, the size of the experiment made affordable the manual analysis of mutations to identify equivalent mutants.

Another threat to validity is the representativeness of the injected faults. The results reported in this study are based on typical mutation operators for Simulink models. In particular, we used the FIM tool~\cite{bartocci2022fim} and its mutation operators, extended with additional mutation operator to address lookup tables. 


\textbf{Internal validity.} In our experiments, we considered only FOMs, i.e., faulty Simulink models with only one fault/mutation. Models can have multiple faults/mutations that may influence each other. Hence, the results might differ when tested with multi-fault Simulink models. Nevertheless, since most of the existing research on mutation testing focuses on FOMs of software artifacts~\cite{aichernig2013time,titcheu2020selecting}, we assessed our technique with single-fault models, leaving the study of HOMs for future work.



\textbf{Conclusion validity.} Random variations is the main threat to conclusion validity. We mitigate this threat by making thirty independent runs of \changes{the test generation} algorithm\changes{s}. 


\section{Lessons Learned 
}\label{sec6}
We now discuss the lessons learned \changes{from our experiments.}

\textbf{Lesson 1 - It is challenging to generate PBMT-adequate test suites.} Our study shows how none of the two state-of-the-art test generation strategies for Simulink programs we experimented with achieved high mutation score with PBMT. Indeed, PBMT is more laborious 
than regular \changes{MT}: a test case that can \emph{kill} a mutant might not \emph{$\phi$-kill} the same mutant. The embedded software industry heavily relies on properties for verification and validation activities, and it is important to design testing tools that thoroughly exercise the software. The definition of PBMT is a relevant advance to the state-of-the-practice that may influence and guide the design of more sophisticated and effective test generation strategies. 


\textbf{Lesson 2 - MT does not capture well the thoroughness of a test suite.} MT can still be applied to Simulink programs. However, test generation techniques could easily kill mutants as long as properties are not considered. This reveals that it is important to not only design executions that cover mutants, but that also propagate the errors produced by mutants, amplifying its visibility on the outputs. These characteristics of a test are not well assessed with MT.



\textbf{Lesson 3 - PBMT-driven test case generation can result in effective test cases.} We defined a SBTG technique to find test cases that demonstrate that mutants could be $\phi$-killed. Such a strategy has been highly effective in \changes{$\phi$-}killing mutants and could be the basis for the design of a mutation-based test case generation strategy. 


\textbf{Lesson 4 - Not all mutations are equally useful to test CPS Simulink models.} Based on our results, we might deduce that some operators are more likely to generate $\phi$-trivially different mutants. For instance, the \texttt{Absolute} operator always generated equivalent mutants. On the other hand, some operators (e.g., \texttt{Negate}, \texttt{ROR}, and \texttt{ASR}) generated mutants that were hard to \changes{$\phi$-}kill, calling for test case generation techniques that exercise the software in non-trivial ways.

\section{Related Work}\label{sec8}

\textbf{Mutation Testing.}
From the software engineering perspective, mutation analysis is one of the powerful software testing techniques that can evaluate the test suite quality~\cite{DeMilloLS78,AcreeBDLS79}. The mutation testing and analysis literature includes a large number of theoretical studies and empirical investigations of various kinds of software artifacts~\cite{jia2010analysis,silva2017systematic}.

The work in~\cite{papadakis2011automatically} combines symbolic execution, concolic execution, and evolutionary testing to automate the test generation for weak mutation testing of programs. Along a similar line of research, the work in~\cite{papadakis2012mutation} proposes a path selection strategy to pick up test cases capable of killing the mutants. Related research on test suite minimization include techniques based on Integer Linear Programming (ILP)~\cite{palomo2018test}, Greedy algorithms~\cite{jatana2016test,polo2012reduction}, formal concept analysis~\cite{li2015test}, etc.

The most prominent works concerning the applicability of mutation testing to safety-critical industrial systems include the empirical investigations reported in~\cite{baker2012empirical,ramler2017empirical,delgado2018evaluation,olivares2021mutation}. Although the work in~\cite{olivares2021mutation} proposes a well-defined mutation analysis pipeline for test suite quality assessment of embedded software, it misses to address the importance of properties associated with the software and the ways to handle them during mutation testing. Contrary to the existing research on regular \changes{MT}, we use properties\textemdash which allow us to express software requirements and specifications\textemdash to formalize the notion of killing the mutants. 

\textbf{Mutations with Simulink models.} 
Mutation mainly relies on alterations in the Simulink model by seeding defects using mutation operators~\cite{binh2012mutation}. Researchers have proposed several tools 
for creating mutants: 
SIMULTATE~\cite{Pill}, MODIFI~\cite{Svenningsson}, ErrorSim~\cite{Sarao}, FIBlock~\cite{Fabarisov}, and FIM~\cite{bartocci2022fim}. We also mention SLforge~\cite{chowdhury2018automatically}, a tool for automatically generating random valid Simulink models for differential testing. 
In our experiments, we used FIM since it provides a higher degree of automation compared to the other tools.



\textbf{Mutation-based test case generation}. With regular MT, the mutation-based test case generation approaches exploit the mutants to generate test cases that can pick up the errors and discover the mutants. 
Some approaches considered generating tests that can reveal mutations introduced in the specification (e.g., in UML models)~\cite{krenn2015momut,aichernig2014model,aichernig2015killing,aichernig2011efficient,krenn2016mutation,fellner2019model}. PBMT is different in many ways: it does not target mutations in the specification and it introduces a novel notion of mutation testing.

The approaches designed to address Simulink models 
focus on targeted test-data generation either using search-based testing~\cite{zhan2005search,zhan2008search} or behavioral analysis approaches (for instance, bounded reachability)~\cite{brillout2009mutation,he2011test}. In essence, the main objective of these techniques is to generate a mutation-adequate test suite that achieves full mutation coverage based on the RIP model. Inspired by these techniques, we designed our search strategy to automatically $\phi$-kill mutants. Further, PBMT introduces a novel instance of mutation testing that assesses the mutation adequacy of test suites w.r.t. properties, which has not been considered in mutation-based testing so far.

\section{Conclusion}\label{sec9}
We presented Property-based Mutation Testing (PBMT), a novel approach to mutation testing that promises efficient evaluation of test suites concerning software properties. Our formalization of mutant killability concerns with the satisfaction (and violation) of a property for the original program (and its mutated version). We provide rigorous semantics for PBMT and its associated mutant killing problem, enabling search-based generation of test cases using a global optimizer. 
We used different test generation strategies for creating test suites and observed their impact on mutant killability.

We studied PBMT on two Simulink models across the safety-critical CPS domain, providing evidence that testing software against properties is more challenging and relevant than opting for regular \changes{MT}, in which mutants can be easily killed. Finally, our evaluation shows that state-of-the-art Adaptive Random Testing and Falsification Testing techniques are still weak in terms of their capability of generating test suites that can effectively kill mutants when tested against properties. 

Future work concerns \changes{adapting} PBMT to closely related CPS modeling languages, including Simulink models integrated with Stateflow Charts. We further plan to conduct additional investigations with HOMs. 

\section*{Acknowledgment}
This work has been supported by the Doctoral College Resilient Embedded Systems, which is run jointly by the TU Wien’s Faculty of Informatics and the UAS Technikum Wien.

\bibliographystyle{IEEEtran}
\bibliography{refs}

\begin{thebibliography}{10}
\providecommand{\url}[1]{#1}
\csname url@samestyle\endcsname
\providecommand{\newblock}{\relax}
\providecommand{\bibinfo}[2]{#2}
\providecommand{\BIBentrySTDinterwordspacing}{\spaceskip=0pt\relax}
\providecommand{\BIBentryALTinterwordstretchfactor}{4}
\providecommand{\BIBentryALTinterwordspacing}{\spaceskip=\fontdimen2\font plus
\BIBentryALTinterwordstretchfactor\fontdimen3\font minus
  \fontdimen4\font\relax}
\providecommand{\BIBforeignlanguage}[2]{{%
\expandafter\ifx\csname l@#1\endcsname\relax
\typeout{** WARNING: IEEEtran.bst: No hyphenation pattern has been}%
\typeout{** loaded for the language `#1'. Using the pattern for}%
\typeout{** the default language instead.}%
\else
\language=\csname l@#1\endcsname
\fi
#2}}
\providecommand{\BIBdecl}{\relax}
\BIBdecl

\bibitem{DeMilloLS78}
R.~A. DeMillo, R.~J. Lipton, and F.~G. Sayward, ``Hints on test data selection:
  Help for the practicing programmer,'' \emph{Computer}, vol.~11, no.~4, pp.
  34--41, April 1978.

\bibitem{AcreeBDLS79}
A.~T. Acree, T.~A. Budd, R.~A. DeMillo, R.~J. Lipton, and F.~G. Sayward,
  ``Mutation analysis,'' Georgia Institute of Technology, Atlanta, Georgia,
  techreport GIT-ICS-79/08, 1979.

\bibitem{olivares2021mutation}
O.~E.~C. Olivares, F.~Pastore, and L.~Briand, ``Mutation analysis for
  cyber-physical systems: Scalable solutions and results in the space domain,''
  \emph{IEEE Transactions on Software Engineering}, 2021.

\bibitem{fortunato2022mutation}
D.~Fortunato, J.~Campos, and R.~Abreu, ``Mutation testing of quantum programs:
  A case study with qiskit,'' \emph{IEEE Transactions on Quantum Engineering},
  vol.~3, pp. 1--17, 2022.

\bibitem{maler2013monitoring}
O.~Maler and D.~Ni{\v{c}}kovi{\'c}, ``Monitoring properties of analog and
  mixed-signal circuits,'' \emph{International Journal on Software Tools for
  Technology Transfer}, vol.~15, no.~3, pp. 247--268, 2013.

\bibitem{7927962}
R.~Gopinath, C.~Jensen, and A.~Groce, ``The theory of composite faults,'' in
  \emph{2017 IEEE International Conference on Software Testing, Verification
  and Validation (ICST)}, 2017, pp. 47--57.

\bibitem{Petrovic:Mutants:ICSE:2021}
G.~Petrovic, M.~Ivankovic, G.~Fraser, and R.~Just, ``Does mutation testing
  improve testing practices?'' in \emph{Proceedings of the International
  Conference on Software Engineering (ICSE)}, 2021.

\bibitem{Just:Mutants:FSE:2014}
R.~Just, D.~Jalali, L.~Inozemtseva, M.~D. Ernst, R.~Holmes, and G.~Fraser,
  ``Are mutants a valid substitute for real faults in software testing?'' in
  \emph{Proceedings of the Symposium on the Foundations of Software Engineering
  (FSE)}, 2014.

\bibitem{Andrews:Mutantion:TSE:2006}
J.~H. Andrews, L.~C. Briand, Y.~Labiche, and A.~S. Namin, ``Using mutation
  analysis for assessing and comparing testing coverage criteria,'' \emph{IEEE
  Transactions on Software Engineering}, vol.~32, no.~8, pp. 608--624, 2006.

\bibitem{Andrews:Mutants:ICSE:2005}
J.~H. Andrews, L.~C. Briand, and Y.~Labiche, ``Is mutation an appropriate tool
  for testing experiments?'' in \emph{Proceedings of the International
  Conference on Software Engineering (ICSE)}, 2015.

\bibitem{jia2009higher}
Y.~Jia and M.~Harman, ``Higher order mutation testing,'' \emph{Information and
  Software Technology}, vol.~51, no.~10, pp. 1379--1393, 2009.

\bibitem{demeyer2020formal}
S.~Demeyer, A.~Parsai, S.~Vercammen, B.~v. Bladel, and M.~Abdi, ``Formal
  verification of developer tests: a research agenda inspired by mutation
  testing,'' in \emph{International Symposium on Leveraging Applications of
  Formal Methods}.\hskip 1em plus 0.5em minus 0.4em\relax Springer, 2020, pp.
  9--24.

\bibitem{vercammen2022mutation}
S.~Vercammen, S.~Demeyer, M.~Borg, N.~Pettersson, and G.~Hedin, ``Mutation
  testing optimisations using the clang front-end,'' \emph{arXiv preprint
  arXiv:2210.17215}, 2022.

\bibitem{papadakis2019mutation}
M.~Papadakis, M.~Kintis, J.~Zhang, Y.~Jia, Y.~Le~Traon, and M.~Harman,
  ``Mutation testing advances: an analysis and survey,'' in \emph{Advances in
  Computers}.\hskip 1em plus 0.5em minus 0.4em\relax Elsevier, 2019, vol. 112,
  pp. 275--378.

\bibitem{kurtz2015static}
B.~Kurtz, P.~Ammann, and J.~Offutt, ``Static analysis of mutant subsumption,''
  in \emph{2015 IEEE Eighth International Conference on Software Testing,
  Verification and Validation Workshops (ICSTW)}.\hskip 1em plus 0.5em minus
  0.4em\relax IEEE, 2015, pp. 1--10.

\bibitem{offutt1997automatically}
A.~J. Offutt and J.~Pan, ``Automatically detecting equivalent mutants and
  infeasible paths,'' \emph{Software testing, verification and reliability},
  vol.~7, no.~3, pp. 165--192, 1997.

\bibitem{offutt2001mutation}
A.~J. Offutt and R.~H. Untch, ``Mutation 2000: Uniting the orthogonal,''
  \emph{Mutation testing for the new century}, pp. 34--44, 2001.

\bibitem{li2016test}
N.~Li and J.~Offutt, ``Test oracle strategies for model-based testing,''
  \emph{IEEE Transactions on Software Engineering}, vol.~43, no.~4, pp.
  372--395, 2016.

\bibitem{chowdhury2018automatically}
S.~A. Chowdhury, S.~Mohian, S.~Mehra, S.~Gawsane, T.~T. Johnson, and
  C.~Csallner, ``Automatically finding bugs in a commercial cyber-physical
  system development tool chain with slforge,'' in \emph{Proceedings of the
  40th International Conference on Software Engineering}, 2018, pp. 981--992.

\bibitem{chowdhury2018curated}
S.~A. Chowdhury, L.~S. Varghese, S.~Mohian, T.~T. Johnson, and C.~Csallner, ``A
  curated corpus of simulink models for model-based empirical studies,'' in
  \emph{2018 IEEE/ACM 4th International Workshop on Software Engineering for
  Smart Cyber-Physical Systems (SEsCPS)}.\hskip 1em plus 0.5em minus
  0.4em\relax IEEE, 2018, pp. 45--48.

\bibitem{rajhans2018graphical}
A.~Rajhans, S.~Avadhanula, A.~Chutinan, P.~J. Mosterman, and F.~Zhang,
  ``Graphical modeling of hybrid dynamics with simulink and stateflow,'' in
  \emph{Proceedings of the 21st International Conference on Hybrid Systems:
  Computation and Control (part of CPS Week)}, 2018, pp. 247--252.

\bibitem{maler2004monitoring}
O.~Maler and D.~Nickovic, ``Monitoring temporal properties of continuous
  signals,'' in \emph{Formal Techniques, Modelling and Analysis of Timed and
  Fault-Tolerant Systems}.\hskip 1em plus 0.5em minus 0.4em\relax Springer,
  2004, pp. 152--166.

\bibitem{donze2010robust}
A.~Donz{\'e} and O.~Maler, ``Robust satisfaction of temporal logic over
  real-valued signals,'' in \emph{International Conference on Formal Modeling
  and Analysis of Timed Systems}.\hskip 1em plus 0.5em minus 0.4em\relax
  Springer, 2010, pp. 92--106.

\bibitem{Simulink}
\BIBentryALTinterwordspacing
Mathworks. (2022) Simulink — simulation and model-based design. [Online].
  Available: \url{https://in.mathworks.com/products/simulink.html}
\BIBentrySTDinterwordspacing

\bibitem{MBSEng}
\BIBentryALTinterwordspacing
Mathworks. (2022) Model-based systems engineering (mbse). [Online]. Available:
  \url{https://in.mathworks.com/solutions/model-based-systems-engineering.html}
\BIBentrySTDinterwordspacing

\bibitem{yadav2021blood}
D.~Yadav, ``Blood coagulation algorithm: A novel bio-inspired meta-heuristic
  algorithm for global optimization,'' \emph{Mathematics}, vol.~9, no.~23, p.
  3011, 2021.

\bibitem{coello2002theoretical}
C.~A.~C. Coello, ``Theoretical and numerical constraint-handling techniques
  used with evolutionary algorithms: a survey of the state of the art,''
  \emph{Computer methods in applied mechanics and engineering}, vol. 191, no.
  11-12, pp. 1245--1287, 2002.

\bibitem{polo2012reduction}
M.~Polo~Usaola, P.~Reales~Mateo, and B.~P{\'e}rez~Lamancha, ``Reduction of test
  suites using mutation,'' in \emph{International Conference on Fundamental
  Approaches to Software Engineering}.\hskip 1em plus 0.5em minus 0.4em\relax
  Springer, 2012, pp. 425--438.

\bibitem{DBLP:conf/atva/Nickovic020}
D.~Nickovic and T.~Yamaguchi, ``{RTAMT:} online robustness monitors from
  {STL},'' in \emph{Automated Technology for Verification and Analysis - 18th
  International Symposium, {ATVA} 2020, Hanoi, Vietnam, October 19-23, 2020,
  Proceedings}, 2020, pp. 564--571.

\bibitem{ATCS}
\BIBentryALTinterwordspacing
Mathworks. (2022) Modeling an automatic transmission controller. [Online].
  Available:
  \url{https://in.mathworks.com/help/simulink/slref/modeling-an-automatic-transmission-controller.html}
\BIBentrySTDinterwordspacing

\bibitem{AECS}
\BIBentryALTinterwordspacing
Mathworks. (2022) Detect faults in aircraft elevator control system. [Online].
  Available:
  \url{https://in.mathworks.com/help/stateflow/ug/fault-detection-control-logic-in-an-aircraft-elevator-control-system.html}
\BIBentrySTDinterwordspacing

\bibitem{donze2010breach}
A.~Donz{\'e}, ``Breach, a toolbox for verification and parameter synthesis of
  hybrid systems,'' in \emph{International Conference on Computer Aided
  Verification}.\hskip 1em plus 0.5em minus 0.4em\relax Springer, 2010, pp.
  167--170.

\bibitem{bartocci2021cpsdebug}
E.~Bartocci, N.~Manjunath, L.~Mariani, C.~Mateis, and D.~Ni{\v{c}}kovi{\'c},
  ``{C}{P}{S}debug: Automatic failure explanation in {C}{P}{S} models,''
  \emph{International Journal on Software Tools for Technology Transfer},
  vol.~23, no.~5, pp. 783--796, 2021.

\bibitem{bartocci2022fim}
E.~Bartocci, L.~Mariani, D.~Ni{\v{c}}kovi{\'c}, and D.~Yadav, ``{FIM:} {F}ault
  {I}njection and {M}utation for {S}imulink,'' in \emph{Proceedings of the 30th
  ACM Joint European Software Engineering Conference and Symposium on the
  Foundations of Software Engineering (ESEC/FSE)}, 2022, pp. 1716--1720.

\bibitem{liu2019effective}
B.~Liu, S.~Nejati, L.~C. Briand \emph{et~al.}, ``Effective fault localization
  of automotive simulink models: achieving the trade-off between test oracle
  effort and fault localization accuracy,'' \emph{Empirical Software
  Engineering}, vol.~24, no.~1, pp. 444--490, 2019.

\bibitem{DBLP:conf/cav/AdimoolamDDKJ17}
A.~S. Adimoolam, T.~Dang, A.~Donz{\'{e}}, J.~Kapinski, and X.~Jin,
  ``Classification and coverage-based falsification for embedded control
  systems,'' in \emph{Computer Aided Verification - 29th International
  Conference, {CAV} 2017, Heidelberg, Germany, July 24-28, 2017, Proceedings,
  Part {I}}, ser. Lecture Notes in Computer Science, R.~Majumdar and V.~Kuncak,
  Eds., vol. 10426.\hskip 1em plus 0.5em minus 0.4em\relax Cham: Springer,
  2017, pp. 483--503.

\bibitem{DBLP:conf/cav/AkazakiH15}
T.~Akazaki and I.~Hasuo, ``Time robustness in {MTL} and expressivity in hybrid
  system falsification,'' in \emph{Computer Aided Verification - 27th
  International Conference, {CAV} 2015, San Francisco, CA, USA, July 18-24,
  2015, Proceedings, Part {II}}, ser. Lecture Notes in Computer Science,
  D.~Kroening and C.~S. Pasareanu, Eds., vol. 9207.\hskip 1em plus 0.5em minus
  0.4em\relax Cham: Springer, 2015, pp. 356--374.

\bibitem{DBLP:conf/cav/ZhangHA19}
Z.~Zhang, I.~Hasuo, and P.~Arcaini, ``Multi-armed bandits for boolean
  connectives in hybrid system falsification,'' in \emph{Computer Aided
  Verification - 31st International Conference, {CAV} 2019, New York City, NY,
  USA, July 15-18, 2019, Proceedings, Part {I}}, ser. Lecture Notes in Computer
  Science, I.~Dillig and S.~Tasiran, Eds., vol. 11561.\hskip 1em plus 0.5em
  minus 0.4em\relax Cham: Springer, 2019, pp. 401--420.

\bibitem{DBLP:conf/cav/ZhangLAMHZ21}
Z.~Zhang, D.~Lyu, P.~Arcaini, L.~Ma, I.~Hasuo, and J.~Zhao, ``Effective hybrid
  system falsification using monte carlo tree search guided by qb-robustness,''
  in \emph{Computer Aided Verification - 33rd International Conference, {CAV}
  2021, Virtual Event, July 20-23, 2021, Proceedings, Part {I}}, ser. Lecture
  Notes in Computer Science, A.~Silva and K.~R.~M. Leino, Eds., vol.
  12759.\hskip 1em plus 0.5em minus 0.4em\relax Cham: Springer, 2021, pp.
  595--618.

\bibitem{aichernig2013time}
B.~K. Aichernig, F.~Lorber, and D.~Ni{\v{c}}kovi{\'c}, ``Time for
  mutants—model-based mutation testing with timed automata,'' in
  \emph{International Conference on Tests and Proofs}.\hskip 1em plus 0.5em
  minus 0.4em\relax Springer, 2013, pp. 20--38.

\bibitem{titcheu2020selecting}
T.~Titcheu~Chekam, M.~Papadakis, T.~F. Bissyand{\'e}, Y.~Le~Traon, and K.~Sen,
  ``Selecting fault revealing mutants,'' \emph{Empirical Software Engineering},
  vol.~25, no.~1, pp. 434--487, 2020.

\bibitem{jia2010analysis}
Y.~Jia and M.~Harman, ``An analysis and survey of the development of mutation
  testing,'' \emph{IEEE transactions on software engineering}, vol.~37, no.~5,
  pp. 649--678, 2010.

\bibitem{silva2017systematic}
R.~A. Silva, S.~d. R.~S. de~Souza, and P.~S.~L. de~Souza, ``A systematic review
  on search based mutation testing,'' \emph{Information and Software
  Technology}, vol.~81, pp. 19--35, 2017.

\bibitem{papadakis2011automatically}
M.~Papadakis and N.~Malevris, ``Automatically performing weak mutation with the
  aid of symbolic execution, concolic testing and search-based testing,''
  \emph{Software Quality Journal}, vol.~19, no.~4, pp. 691--723, 2011.

\bibitem{papadakis2012mutation}
M.~Papadakis and N.~Malevris, ``Mutation based test case generation via a path
  selection strategy,'' \emph{Information and Software Technology}, vol.~54,
  no.~9, pp. 915--932, 2012.

\bibitem{palomo2018test}
F.~Palomo-Lozano, A.~Estero-Botaro, I.~Medina-Bulo, and M.~N{\'u}{\~n}ez,
  ``Test suite minimization for mutation testing of ws-bpel compositions,'' in
  \emph{Proceedings of the Genetic and Evolutionary Computation Conference},
  2018, pp. 1427--1434.

\bibitem{jatana2016test}
N.~Jatana, B.~Suri, P.~Kumar, and B.~Wadhwa, ``Test suite reduction by mutation
  testing mapped to set cover problem,'' in \emph{Proceedings of the Second
  International Conference on Information and Communication Technology for
  Competitive Strategies}, 2016, pp. 1--6.

\bibitem{li2015test}
L.~Li and H.~Gao, ``Test suite reduction for mutation testing based on formal
  concept analysis,'' in \emph{2015 IEEE/ACIS 16th International Conference on
  Software Engineering, Artificial Intelligence, Networking and
  Parallel/Distributed Computing (SNPD)}.\hskip 1em plus 0.5em minus
  0.4em\relax IEEE, 2015, pp. 1--5.

\bibitem{baker2012empirical}
R.~Baker and I.~Habli, ``An empirical evaluation of mutation testing for
  improving the test quality of safety-critical software,'' \emph{IEEE
  Transactions on Software Engineering}, vol.~39, no.~6, pp. 787--805, 2012.

\bibitem{ramler2017empirical}
R.~Ramler, T.~Wetzlmaier, and C.~Klammer, ``An empirical study on the
  application of mutation testing for a safety-critical industrial software
  system,'' in \emph{Proceedings of the Symposium on Applied Computing}, 2017,
  pp. 1401--1408.

\bibitem{delgado2018evaluation}
P.~Delgado-P{\'e}rez, I.~Habli, S.~Gregory, R.~Alexander, J.~Clark, and
  I.~Medina-Bulo, ``Evaluation of mutation testing in a nuclear industry case
  study,'' \emph{IEEE Transactions on Reliability}, vol.~67, no.~4, pp.
  1406--1419, 2018.

\bibitem{binh2012mutation}
N.~T. Binh \emph{et~al.}, ``Mutation operators for simulink models,'' in
  \emph{2012 Fourth International Conference on Knowledge and Systems
  Engineering}.\hskip 1em plus 0.5em minus 0.4em\relax IEEE, 2012, pp. 54--59.

\bibitem{Pill}
I.~Pill, I.~Rubil, F.~Wotawa, and M.~Nica, ``Simultate: A toolset for fault
  injection and mutation testing of simulink models,'' in \emph{2016 IEEE Ninth
  International Conference on Software Testing, Verification and Validation
  Workshops (ICSTW)}.\hskip 1em plus 0.5em minus 0.4em\relax IEEE, 2016, pp.
  168--173.

\bibitem{Svenningsson}
R.~Svenningsson, J.~Vinter, H.~Eriksson, and M.~T{\"o}rngren, ``Modifi: a
  model-implemented fault injection tool,'' in \emph{International Conference
  on Computer Safety, Reliability, and Security}, E.~Schoitsch, Ed.\hskip 1em
  plus 0.5em minus 0.4em\relax Berlin, Heidelberg: Springer, 2010, pp.
  210--222.

\bibitem{Sarao}
M.~Sarao{\u{g}}lu, A.~Morozov, M.~T. S{\"o}ylemez, and K.~Janschek, ``Errorsim:
  A tool for error propagation analysis of simulink models,'' in
  \emph{International Conference on Computer Safety, Reliability, and
  Security}.\hskip 1em plus 0.5em minus 0.4em\relax Springer, 2017, pp.
  245--254.

\bibitem{Fabarisov}
\BIBentryALTinterwordspacing
T.~Fabarisov, I.~Mamaev, A.~Morozov, and K.~Janschek, ``Model-based fault
  injection experiments for the safety analysis of exoskeleton system,''
  \emph{arXiv preprint arXiv:2101.01283}, 2021. [Online]. Available:
  \url{https://arxiv.org/abs/2101.01283}
\BIBentrySTDinterwordspacing

\bibitem{krenn2015momut}
W.~Krenn, R.~Schlick, S.~Tiran, B.~Aichernig, E.~Jobstl, and H.~Brandl,
  ``Momut:: Uml model-based mutation testing for uml,'' in \emph{2015 IEEE 8th
  International Conference on Software Testing, Verification and Validation
  (ICST)}.\hskip 1em plus 0.5em minus 0.4em\relax IEEE, 2015, pp. 1--8.

\bibitem{aichernig2014model}
B.~K. Aichernig, J.~Auer, E.~J{\"o}bstl, R.~Koro{\v{s}}ec, W.~Krenn,
  R.~Schlick, and B.~V. Schmidt, ``Model-based mutation testing of an
  industrial measurement device,'' in \emph{International Conference on Tests
  and Proofs}.\hskip 1em plus 0.5em minus 0.4em\relax Springer, 2014, pp.
  1--19.

\bibitem{aichernig2015killing}
B.~K. Aichernig, H.~Brandl, E.~J{\"o}bstl, W.~Krenn, R.~Schlick, and S.~Tiran,
  ``Killing strategies for model-based mutation testing,'' \emph{Software
  Testing, Verification and Reliability}, vol.~25, no.~8, pp. 716--748, 2015.

\bibitem{aichernig2011efficient}
B.~K. Aichernig, H.~Brandl, E.~J{\"o}bstl, and W.~Krenn, ``Efficient mutation
  killers in action,'' in \emph{2011 Fourth IEEE International Conference on
  Software Testing, Verification and Validation}.\hskip 1em plus 0.5em minus
  0.4em\relax IEEE, 2011, pp. 120--129.

\bibitem{krenn2016mutation}
W.~Krenn and R.~Schlick, ``Mutation-driven test case generation using
  short-lived concurrent mutants--first results,'' \emph{arXiv preprint
  arXiv:1601.06974}, 2016.

\bibitem{fellner2019model}
A.~Fellner, W.~Krenn, R.~Schlick, T.~Tarrach, and G.~Weissenbacher,
  ``Model-based, mutation-driven test-case generation via heuristic-guided
  branching search,'' \emph{ACM Transactions on Embedded Computing Systems
  (TECS)}, vol.~18, no.~1, pp. 1--28, 2019.

\bibitem{zhan2005search}
Y.~Zhan and J.~A. Clark, ``Search-based mutation testing for simulink models,''
  in \emph{Proceedings of the 7th annual conference on Genetic and evolutionary
  computation}, 2005, pp. 1061--1068.

\bibitem{zhan2008search}
Y.~Zhan and J.~A. Clark, ``A search-based framework for automatic testing of
  matlab/simulink models,'' \emph{Journal of Systems and Software}, vol.~81,
  no.~2, pp. 262--285, 2008.

\bibitem{brillout2009mutation}
A.~Brillout, N.~He, M.~Mazzucchi, D.~Kroening, M.~Purandare, P.~R{\"u}mmer, and
  G.~Weissenbacher, ``Mutation-based test case generation for simulink
  models,'' in \emph{International Symposium on Formal Methods for Components
  and Objects}.\hskip 1em plus 0.5em minus 0.4em\relax Springer, 2009, pp.
  208--227.

\bibitem{he2011test}
N.~He, P.~R{\"u}mmer, and D.~Kroening, ``Test-case generation for embedded
  simulink via formal concept analysis,'' in \emph{Proceedings of the 48th
  Design Automation Conference}, 2011, pp. 224--229.

\end{thebibliography}

\end{document}